\documentclass[graybox]{svmult}

\usepackage{mathptmx}       
\usepackage{helvet}         
\usepackage{courier}        
\usepackage{type1cm}        
\usepackage{makeidx}         
\usepackage{graphicx}        
\usepackage{multicol}        
\usepackage[bottom]{footmisc}

\usepackage{amsmath}
\usepackage[numbers]{natbib}
\usepackage{todonotes}
\newcommand{\bq}{\begin{equation}}
\newcommand{\eq}{\end{equation}}

\newcommand{\flop}{\mbox{flop}}
\newcommand{\flops}{\mbox{flops}}

\newcommand{\GFS}{\mbox{Gflop/s}}

\newcommand{\MLUPS}{\mbox{MLUP/s}}

\newcommand{\bytes}{\mbox{bytes}}

\newcommand{\cycle}{\mbox{cy}}

\newcommand{\eos}{~.}

\newcommand{\construction}[1]{}

\newcommand{\olsep}{\|}
\newcommand{\nolsep}{|}
\newcommand{\ecmspace}{\,}
\newcommand{\ecm}[6]{\mbox{$\left\{{#1}\ecmspace\olsep\ecmspace {#2}\ecmspace\nolsep\ecmspace {#3}\ecmspace\nolsep\ecmspace {#4}\ecmspace\nolsep\ecmspace {#5}\right\}\ecmspace{#6}$}}

\usepackage{multirow}
\usepackage{hyperref}
\usepackage[final]{listings}
\definecolor{MidnightBlue}{cmyk}{0.97,0.78,0.39,0.29}
\definecolor{MidnightBlue}{RGB}{25,25,112}

\lstdefinelanguage{shell}
{
sensitive=false,
morestring=[b]",
morecomment=[s][\color{orange}]{$}{;},
}

\newcommand\YAMLkeystyle{\color{black}\bfseries\ttfamily}
\newcommand\YAMLvaluestyle{\color{black}\mdseries\ttfamily}

\lstdefinelanguage{yaml}
{
  basicstyle=\YAMLkeystyle,                                 
  sensitive=false,
  comment=[l]{\#},
  morecomment=[s]{/*}{*/},
  commentstyle=\color{orange}\ttfamily,
  moredelim=[l][\color{orange}]{\&},
  moredelim=[l][\color{magenta}]{*},
  moredelim=**[il][\YAMLkeystyle{:}\YAMLvaluestyle]{:},   
  morestring=[b]',
  morestring=[b]",
  literate =    {---}{{\ProcessThreeDashes}}3
                {>}{{\textcolor{red}\textgreater}}1     
                {|}{{\textcolor{red}\textbar}}1 
                {\ -\ }{{\mdseries\ -\ }}3,
}

\newcommand\ProcessThreeDashes{\llap{\color{cyan}\mdseries-{-}-}}

\makeindex             

\begin{document}

\title*{Kerncraft: A Tool for Analytic Performance Modeling of Loop Kernels}
\author{Julian Hammer, Jan Eitzinger, Georg Hager, and Gerhard Wellein}
\authorrunning{J. Hammer, J. Eitzinger, G. Hager, and G. Wellein}
\institute{Julian Hammer \at Erlangen Regional Computing Center, Germany, \email{julian.hammer@fau.de}
\and Jan Eitzinger \at Erlangen Regional Computing Center, Germany, \email{jan.eitzinger@fau.de} 
\and Georg Hager \at Erlangen Regional Computing Center, Germany, \email{georg.hager@fau.de} 
\and Gerhard Wellein \at Erlangen Regional Computing Center, Germany, \email{gerhard.wellein@fau.de}}
%
%


\maketitle

\abstract*{Achieving optimal program performance requires deep insight into
the interaction between hardware and software. For software developers
without an in-depth background in computer architecture, understanding
and fully utilizing modern architectures is close to impossible.
Analytic loop performance modeling is a useful way to understand
the relevant bottlenecks of code execution based on simple machine
models. The Roof{}line Model and the Execution-Cache-Memory (ECM)
model  are proven approaches to performance modeling
of loop nests. In comparison to the Roof{}line model, the ECM model
can also describes the single-core performance and saturation behavior
on a multicore chip.\\
We give an introduction to the Roof{}line and ECM models, and to
stencil performance modeling using layer conditions (LC). We then
present Kerncraft, a tool that can automatically construct
Roof{}line and ECM models for loop nests by performing the
required code, data transfer, and LC analysis. The layer condition
analysis allows to predict optimal spatial blocking factors
for loop nests. Together with the models it enables an ab-initio
estimate of the 
potential benefits of loop blocking optimizations and of useful block sizes.
In cases where LC analysis is not easily possible, Kerncraft 
supports a cache simulator as a fallback option.
Using a 25-point long-range stencil we demonstrate the usefulness and
predictive power of the Kerncraft tool.}

\abstract{Achieving optimal program performance requires deep insight into
the interaction between hardware and software. For software developers
without an in-depth background in computer architecture, understanding
and fully utilizing modern architectures is close to impossible.
Analytic loop performance modeling is a useful way to understand
the relevant bottlenecks of code execution based on simple machine
models. The Roof{}line Model and the Execution-Cache-Memory (ECM)
model  are proven approaches to performance modeling
of loop nests. In comparison to the Roof{}line model, the ECM model
can also describes the single-core performance and saturation behavior
on a multicore chip.\\
We give an introduction to the Roof{}line and ECM models, and to
stencil performance modeling using layer conditions (LC). We then
present Kerncraft, a tool that can automatically construct
Roof{}line and ECM models for loop nests by performing the
required code, data transfer, and LC analysis. The layer condition
analysis allows to predict optimal spatial blocking factors
for loop nests. Together with the models it enables an ab-initio
estimate of the 
potential benefits of loop blocking optimizations and of useful block sizes.
In cases where LC analysis is not easily possible, Kerncraft 
supports a cache simulator as a fallback option.
Using a 25-point long-range stencil we demonstrate the usefulness and
predictive power of the Kerncraft tool.}


\section{Introduction}
\label{sec:introduction}

Expensive, large-scale supercomputers consisting of thousands of nodes make performance 
a major issue for efficient resource utilization. A lot of research in this area concentrates on massive scalability, but there is just as much potential for optimization at the core and chip levels.
If performance fails to be acceptable at small scales, scaling up will waste resources even if the parallel efficiency is good. Therefore, performance engineering should always start with solid insight at the smallest scale: the core. Using this approach will give the performance engineer a profound understanding of performance behavior, guide optimization attempts and, finally, drive scaling at the relevant hardware bottlenecks.

Modeling techniques are essential to understand performance on a single core due to the complexities hidden in modern CPU and node architectures. Without a model it is hardly possible to navigate through the multitude of potential performance bottlenecks such as memory bandwidth, execution unit throughput, decoder throughput, cache latency, TLB misses or even OS jitter, which may or may not be relevant to the specific application at hand. Analytic models, if applied correctly, help us focus on the most relevant factors and allow validation of the gained insights. With ``analytic'' we mean models that were derived not by automated fitting of parameters of a highly generic predictor function, but by consciously selecting key factors that can be explained and understood by experts and then constructing a simplified machine and execution model from them.

We understand that the application of analytic performance modeling techniques often poses challenges or tends to be tedious, even for experienced software developers with a deep understanding of computer architecture and performance engineering. Kerncraft \cite{hammer15kerncraft}, our tool for automatic performance modeling, addresses these issues. Since its first publication, Kerncraft has been throughly extended with the layer condition model, an independent and more versatile cache simulation, as well as more flexible support for data accesses and kernel codes. These enhancements will be detailed in the following sections. Kerncraft is available for download under GPLv3 \cite{kerncraft-github}.

\subsection{Related Work}
\label{subsec:related}

Out of the many performance modeling tools that rely on hardware
metrics, statistical methods, curve fitting, and machine learning,
there are only four projects in the area of automatic and
analytic modeling that we know of: PBound, ExaSAT, Roof\/line Model
Toolkit and MAQAO.

Narayanan et al.\ \cite{pbounds2010} describe a tool (PBound) for
automatically extracting relevant information about execution
resources (arithmetic operations, loads and stores) from source
code. They do not, however, consider cache effects and parallel
execution, and their machine model is rather idealized.  Unat et
al.\ \cite{Unat01052015} introduce the ExaSAT tool, which uses
compiler technology for source code analysis and also employs ``layer
conditions'' \cite{stengel14} to assess the real data traffic for
every memory hierarchy level based on cache and problem sizes. They
use an abstract simplified machine model, whereas our Kerncraft tool
employs Intel IACA  to generate more
accurate in-core predictions. On the one hand this (currently)
restricts Kerncraft's in-core predictions to Intel CPUs, but on the
other hand provides predictions from the actual machine code
containing all compiler optimizationsr. Furthermore, ExaSAT is restricted to the Roof\/line model
for performance prediction. Being compiler-based,
ExaSAT supports full-application modeling and code optimizations,
which is work in progress for Kerncraft. It can
also incorporate communication (i.e., message passing) overhead, 
which is not the scope of our research.  Lo et al.~\cite{rmt15} introduced in 2014 the ``Empirical Roof\/line
Toolkit,'' (ERT) which aims at automatically generating hardware
descriptions for Roof\/line analysis. They do not support automatic
topology detection and their use of compiler-generated loops
introduces an element of uncertainty.  Djoudi et
al.~\cite{djoudi2005maqao} started the MAQAO Project in 2005, which
uses static analysis to predict in-core execution time and combines it
with dynamic analysis to assess the overall code quality. It was
originally developed for the Itanium~2 processor but has since been adapted for
recent Intel64 architectures and the Xeon Phi. As with Kerncraft, MAQAO
currently supports only Intel architectures. The memory access analysis is based on
dynamic run-time data, i.e., it requires the code to be run on
the target architecture.

\subsection{Performance Models}
\label{subsec:perfmodels}

Performance modeling, historically done by pen, paper and brain, has a long tradition in computer science. For instance, the well-known Roof{}line model has its origins in the 1980s \cite{hockney89}. In this paper, we make use of the Roof{}line and the Execution-Cache-Memory (ECM) models, both of which are based on a bottleneck analysis under a throughput assumption. Detailed explanations of the models can be found in previous publications; we will limit ourselves to a short overview.

\subsubsection{Roof{}line}
\label{subsec:roofline}
The Roof\/line model yields an absolute upper performance bound for a loop. It is based on the assumption that either the data transfers to and from a single level in the memory hierarchy or the computational work dictates the runtime. This implies that all data transfers to all memory hierarchy levels perfectly overlap with each other and with the execution of instructions in the core, which is too optimistic in the general case. The Roof{}line model in the current form was popularized and named by Williams et al.\ in 2009 \cite{roofline:2009}.

For the types of analysis Kerncraft supports it is useful to reformulate the Roof\/line model in terms of execution time instead of performance, and to use a basic unit of work that spans the length of a cache line (typically eight iterations): $T_\mathrm{roof}=\max_k\left(T_\mathrm{core},T_k\right)$. The ratio $T_k=\beta_k/B_k$, with the achievable peak bandwidth $B_k$ and data transfer volume $\beta_k$, is the data transfer time for memory hierarchy level $k$. $T_\mathrm{core}=\phi/P_\mathrm{max}$ is the in-core execution time for computations with the amount of work $\phi$. The latter is usually given in \flops, but any other well-defined metric will do. $P_\mathrm{max}$ is the applicable computational peak
performance (in \flops\ per \cycle) of the code at hand. It may be smaller than the
absolute peak performance because of unbalanced multiply and add operations, because SIMD cannot be applied, etc.

Applying the Roof\/line model to a loop kernel which loads 448~\bytes\ from the first level cache (L1), 6 cache lines (CL) from the second level cache (L2), 4 CLs from the last level cache (L3), and two CLs from main memory, to produce one cache line of results (8 iterations), gives us the data volumes in Table~\ref{tbl:roofline}. This is what we would expect with a 3D seven-point stencil (see Listing~\ref{lst:3d-7pt}) for a certain problem size that leads to a 3D-layer condition fulfilled in L3 and a 2D-layer condition fulfilled in L2 (see below for more on layer conditions). For the computational work, we assume 5 additions and 7 multiplications per iteration, thus 96 FLOPs for eight iterations, i.e.,
$\phi = 96\,\flop$.
Combining this with measured bandwidths from a STREAM \cite{STREAM} copy kernel on an Ivy Bridge EP processor in all memory hierarchy levels, we can derive the throughput time per eight iterations shown in the last column of Table~\ref{tbl:roofline}. The achievable peak bandwidth $B_k$ is obtained via a streaming benchmark since theoretical bandwidths published by vendors cannot be obtained in practice.
The ECM model provides a partial explanation for this effect, so it requires less measured input data (see below).
\begin{table}[tbp]
    \caption{\label{tbl:roofline}Overview of data transfers and bandwidths necessary to model a 3D seven-point stencil kernel using the Roof{}line model.}
\centering\renewcommand{\arraystretch}{1.1}
\begin{tabular}{r||r|r|r}
    Level & Data Volume per 8 It. & STREAM copy Bandwidth & Time for 8 It. \\
    $k$ & $\beta_k$ & $B_k$ & $T_k$ \\
    \hline
    \hline
    L1 & $448\,\mbox{B}$ (only LOAD) & $137.1\,\mbox{GB/s}$ & $9.8\,\mbox{cy}$ \\
    L2 & $7\,\mbox{CL}$ or $384\,\mbox{B}$ & $68.4\,\mbox{GB/s}$ & $16.6\,\mbox{cy}$ \\
    L3 & $5\,\mbox{CL}$ or $256\,\mbox{B}$ & $38.8\,\mbox{GB/s}$ & $24.7\,\mbox{cy}$ \\
    MEM & $3\,\mbox{CL}$ or $128\,\mbox{B}$ & $17.9\,\mbox{GB/s}$ & $32.2\,\mbox{cy}$ \\
\end{tabular}
\end{table}

The double precision maximum applicable performance of a code with $5/7$ addition-multiplication ratio on an Ivy Bridge core  is
$$P_{\mathrm{max}} = \frac{40\,\flop}{7\,\cycle}$$
which yields an in-core prediction of
$$T_\mathrm{core}={96\,\flop \over 40\,\flop / 7\,\cycle} = 16.8 \,\mbox{cy}$$
The dominating bottleneck is therefore the transfer from main memory $T_\mathrm{MEM}$ with $32.2\,\mbox{cy}$ for eight iterations or updates, which corresponds to a maximum expected (``lightspeed'') performance of $8.94\,\GFS$.

Predicting the L1 time and performance with the measured bandwidth can only be precise if the microbenchmark mimics exactly the load/store ratio as found in the modeled code. To circumvent this issue it is advisable to use a static analyzer with knowledge of the architecture, like the Intel Architecture Core Analyzer (IACA) \cite{iacaweb}. It also allows a more accurate prediction of $T_{\mathrm{core}}$.

\subsubsection{Execution-Cache-Memory}
\label{subsec:ecm}

The Execution-Memory-Cache (ECM) model is based on the same fundamental idea as the Roof\/line model, i.e., that data transfer time or execution of instructions, whichever takes longer, determine the runtime of a loop.  Unlike in the Roof\/line model, all memory hierarchy levels contribute to a single bottleneck. Depending on the microarchitecture, data transfer times to different memory hierarchy levels may overlap (as in the Roof{}line model) or they may add up. This latter assumption was shown to fit measurements quite well on x86-based processors~\cite{stengel14,Wittmann:2016}; on Power8, for instance, the cache hierarchy shows considerable overlap~\cite{Hofmann:2016-Kahan}. In the following we will concentrate on Intel architectures, since the current version of Kerncraft implements a strict non-overlapping ECM model.

We also need to know the data volumes transferred to and from each memory hierarchy level and the amount of work performed in the core. To calculate the time contributions per cache level we use documented inter-cache throughputs (e.g., two cycles per cache line from L3 to L2 on Intel Ivy Bridge). The ECM prediction on an Intel core for data in memory is then given by 
$$
T_\mathrm{ECM,Mem} = \max\left(T_\mathrm{OL},T_\mathrm{nOL}+T_\mathrm{L1-L2}+
T_\mathrm{L2-L3}+T_\mathrm{L3-MEM}\right)\eos
$$
$T_\mathrm{OL}$ is the overlapping time for computations and stores, $T_\mathrm{nOL}$ is the time for the loads from registers into L1, $T_\mathrm{L1L2}$ the loads from L2 into L1, and so on. The model is  written in the following compact notation:
$$
\ecm{T_\mathrm{OL}}{T_\mathrm{nOL}}{T_\mathrm{L1-L2}}{T_\mathrm{L2-L3}}{T_\mathrm{L3-MEM}}{}\eos
$$
See \cite{stengel14} for more details on the model and the notation.

Applying the ECM model to the 3D seven-point stencil (see Listing~\ref{lst:3d-7pt}) on an Ivy Bridge EP processor, we get the in-core contributions from IACA:
$$T_\mathrm{OL}= 13.2\,\mbox{cy}\quad\mbox{and}\quad
T_\mathrm{nOL} = \beta_\mathrm{L1} \cdot 1 \frac{\mbox{cy}}{64\,\mbox{B}}=7\,\mbox{cy}~~.
$$
The data transfers through the memory hierarchy are obtained from cache simulation in combination with hardware performance characteristics:
$$T_\mathrm{L1-L2} = \beta_\mathrm{L2} \cdot 2\frac{\mbox{cy}}{\mbox{CL}}=14\,\mbox{cy}$$
$$T_\mathrm{L2-L3} = \beta_\mathrm{L3} \cdot 2\frac{\mbox{cy}}{\mbox{CL}}=10\,\mbox{cy}$$
$$T_\mathrm{L3-MEM} = \frac{\beta_\mathrm{MEM} \cdot 3.0\frac{\mbox{Gcy}}{\mbox{s}} \cdot 64\frac{\mbox{B}}{\mbox{CL}}}{63.4 \frac{\mbox{GB}}{\mbox{s}}} = 9.1\,\mbox{cy}$$
The ECM notation for eight iterations of the 3D seven-point stencil code is then:
$$
\ecm{13.2}{7}{14}{10}{9.1}{\mbox{cy}}\eos
$$

\begin{figure}[tbp]
\sidecaption
\includegraphics[scale=.78]{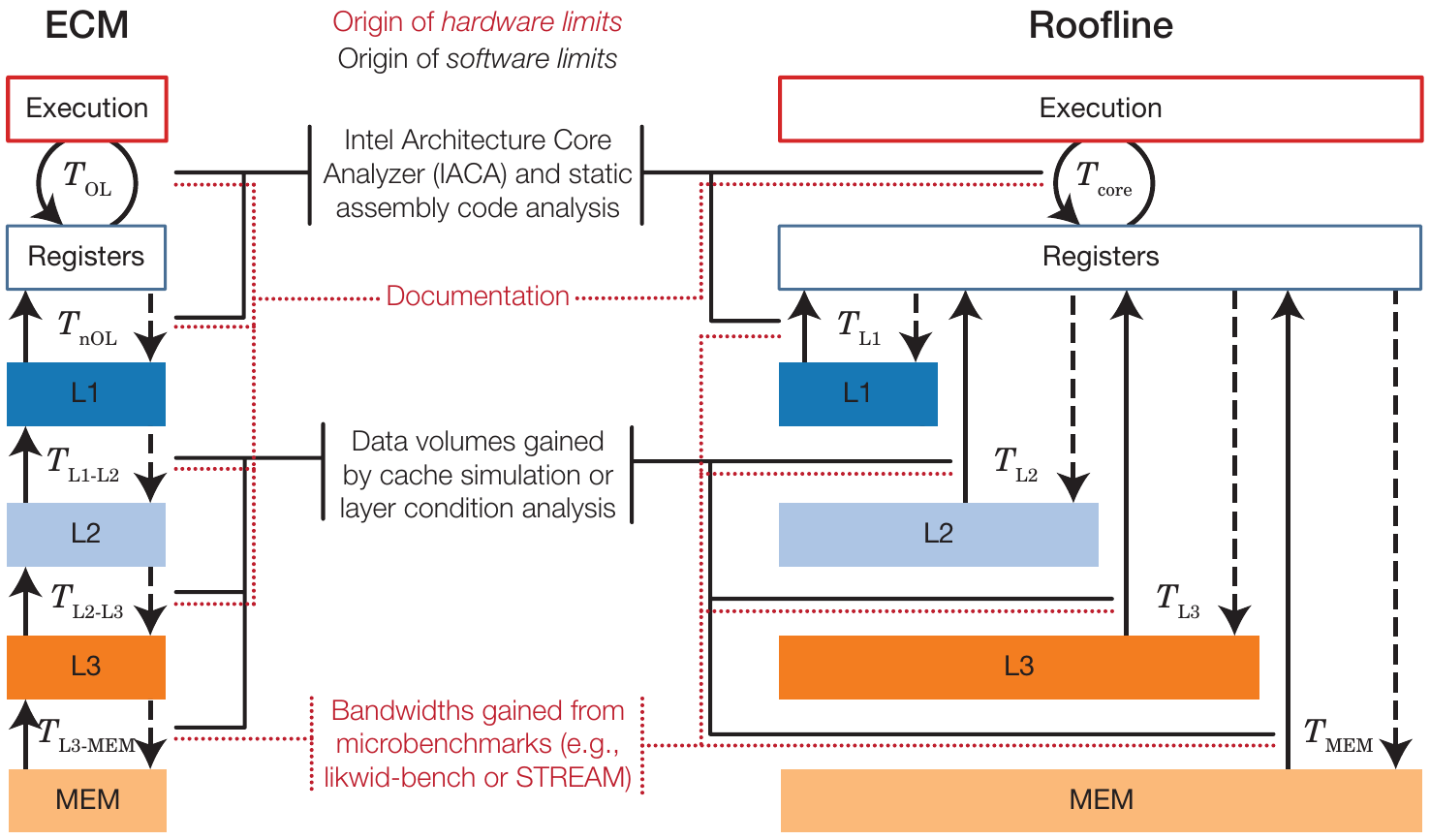}
\caption{Side-by-side comparison of the (x86) ECM model and the Roof{}line model, including the origin of information needed as input for both, such as bandwidth and execution bottlenecks.}
\label{fig:ECMvsRoofline}
\end{figure}

A comparison of the data that goes into the ECM and Roof{}line analysis (manual and automatic) is shown in Figure~\ref{fig:ECMvsRoofline}. It also illustrates the fundamental differences in the bottleneck assumption.

\section{Kerncraft}
\label{sec:kerncraft}
In this section we give an overview of the architecture and analysis modes available in Kerncraft. The recent additions, which have not been covered in our 2015 publication \cite{hammer15kerncraft}, will be explained in due detail.

\begin{figure}[t]
\sidecaption
\includegraphics[width=\textwidth]{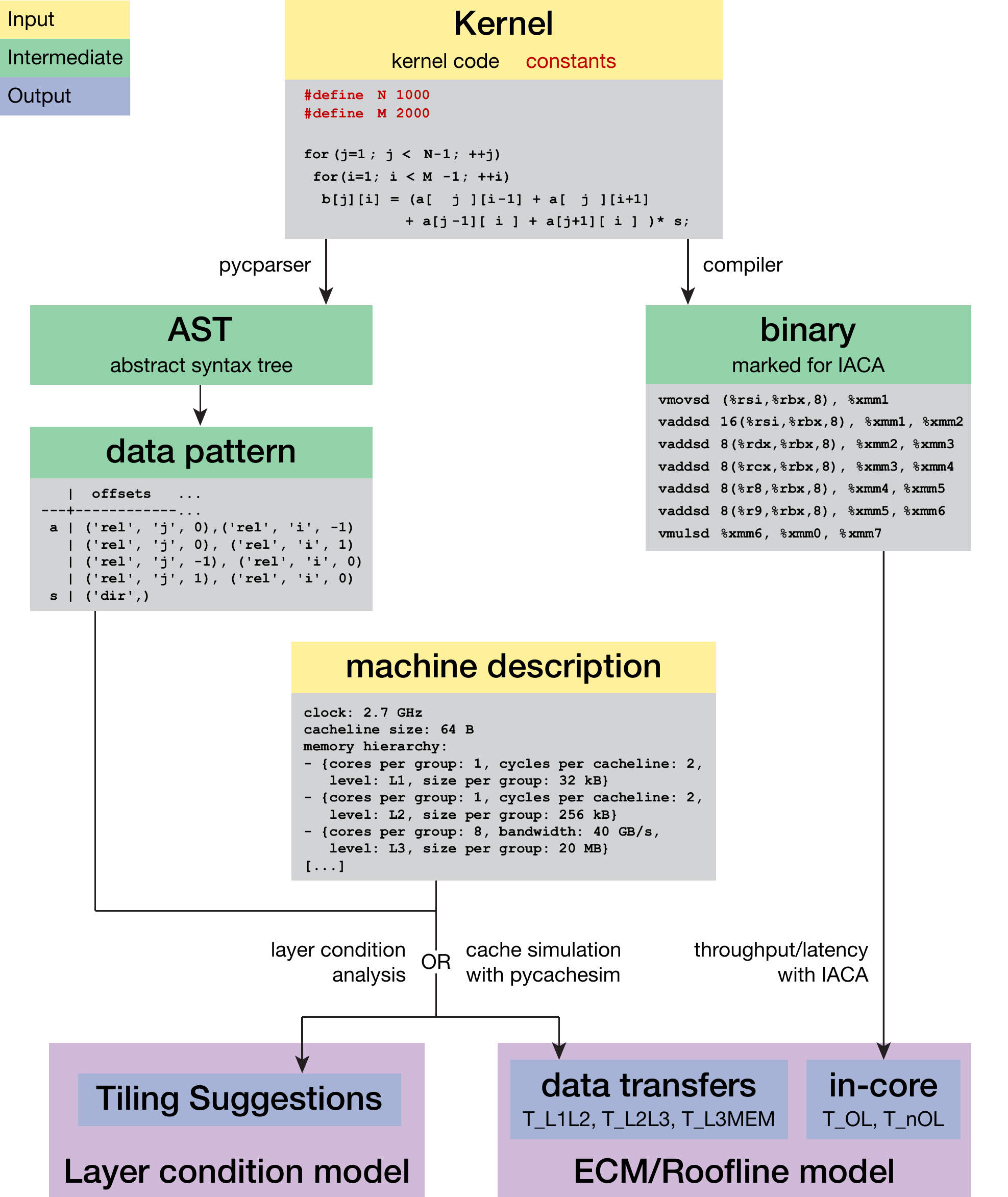}
\caption{Overview of the Kerncraft pipeline. The user provides kernel code, constants, and a machine description. IACA, pycachesim, and a compiler are employed to build the ECM, Roof\/line, and layer condition models.}
\label{fig:overview}
\end{figure}

The core of Kerncraft is responsible for parsing and extracting information from a given kernel code, abstracting information about the machine, and providing a homogenous user interface. The modules responsible for the modeling  will be described in Section~\ref{subsec:models}. A visualization of the overall structure is shown in Figure~\ref{fig:overview}. The user has to provide a kernel code (described in Sec.~\ref{subsec:code}) and a machine description (described in Sec.~\ref{subsec:machine_description}), and they have to select a performance model to apply (options are described in Sec.\ref{subsec:models}). Optionally, parameters can be passed to the kernel code, similar to constants defined by macros or \verb.-D. compiler flags. For models that rely on prediction of caching, either the layer condition prediction or the cache simulation (using the pycachesim module) can be employed. Both predictors will be described in Section~\ref{subsec:cacheprediction}.

\subsection{Kernel Code}
\label{subsec:code}

Kerncraft is based on the analysis of standard-compliant C99 \cite{ISO:C99} code, which must be provided as shown in Listing~\ref{lst:3d-7pt}. Example files for several stencils are distributed with the Kerncraft repository\footnote{\url{https://github.com/RRZE-HPC/kerncraft/tree/master/examples/kernels}}. The first lines are dedicated to variable and array definitions. While large arrays would in practice be allocated on the heap, Kerncraft requires arrays to be declared as local varaibles. The multidimensional syntax (e.g., \texttt{a[M][N]} and \texttt{a[j][i]}) is optional, since Kerncraft now also supports flattened indices (e.g., \texttt{a[M*N]} and \texttt{a[j*N+i]}).

\texttt{N} and \texttt{M} in Listing~\ref{lst:3d-7pt}, are constants which can be passed to the code through the command line. During analysis they are treated as symbols, which may be replaced by constant positive integers.
\begin{lstlisting}[language=C,float=tbp,label={lst:3d-7pt},caption={Input kernel code for a three-dimensional 7-point star stencil (3D-7pt).}]
double a[M][N][N];
double b[M][N][N];
double coeffs_N, coeffs_S, coeffs_W, coeffs_E,
       coeffs_F, coeffs_B, s;

for(int k=1; k<M-1; ++k)
    for(int j=1; j<N-1; ++j)
        for(int i=1; i<N-1; ++i)
            b[k][j][i] = ( coeffs_W*a[k][j][i-1]
                         + coeffs_E*a[k][j][i+1]
                         + coeffs_N*a[k][j-1][i]
                         + coeffs_S*a[k][j+1][i]
                         + coeffs_B*a[k-1][j][i]
                         + coeffs_F*a[k+1][j][i]) * s;
\end{lstlisting}

Following the variable definitions is the loop nest, which may only contain one loop per level and only the innermost loop may contain variable assignments and arithmetic operations.
The loop indices must be local to that loop and the bounds may only depend on constant integers and simple arithmetic operations (addition, subtraction, and multiplication) of constant integers. The step size can be any constant length; in Listing~\ref{lst:3d-7pt} we have a step size of one, but \texttt{k+=2} would for instance also work.

Any number of statements are allowed in the loop body, as long as they are assignments and arithmetic operations based on constants, integers, variables, and array references. Array references may contain arithmetic expressions in their indices (e.g., \texttt{a[j*N+i+1]}). Such an expression may only be composed of constants, integers, and loop index variables (\texttt{i}, \texttt{j}, and \texttt{k} in Listing~\ref{lst:3d-7pt}).

Function calls, \texttt{if}s, pointer arithmetic, and irregular data accesses are not allowed, since they could not be analyzed with the algorithms used in the current version of Kerncraft. Moreover, the underlying models do not yet have a canonical way of dealing with the effects arising in such cases.

\subsection{Machine Description}
\label{subsec:machine_description}
To select the targeted machine architecture, Kerncraft needs a machine description file in the YAML file format \cite{yaml}. Example machines description files are distributed through the Kerncraft repository\footnote{\url{https://github.com/RRZE-HPC/kerncraft/tree/master/examples/machine-files}}. A machine description file always consists of three parts: the execution architecture description, the cache and memory hierarchy description, and benchmark results of typical streaming kernels. In the following, we will go into some settings found in Listing~\ref{lst:machinefile} that are not self-explanatory.

\begin{lstlisting}[float=tbp,language=yaml,label={lst:machinefile},caption={Shortened machine description for Haswell (skipped sections are marked by \texttt{...}).}]
# Execution Architecture:
model name: Intel(R) Xeon(R) CPU E5-2695 v3 @ 2.30GHz
micro-architecture: HSW
non-overlapping ports: [2D, 3D]
overlapping ports: ['0', 0DV, '1', '2', '3', '4', '5', '6', '7']
FLOPs per cycle:
  DP: {ADD: 8, FMA: 8, MUL: 8, total: 16}
  SP: {ADD: 16, FMA: 16, MUL: 16, total: 32}
compiler: icc
compiler flags: [-O3, -xAVX, -fno-alias]
...
# Memory and Cache Hierarchy:
memory hierarchy:
    - level: L1
      cache per group: {
         'sets': 64, 'ways': 8, 'cl_size': 64, # 32 kB
         'replacement_policy': 'LRU',
         'write_allocate': True, 'write_back': True,
         'load_from': 'L2', 'store_to': 'L2'}
      cores per group: 1
      threads per group: 2
      groups: 28
      cycles per cacheline transfer: 2
    ...
# Benchmark Description and Results:
benchmarks:
  kernels:
    copy:
      FLOPs per iteration: 0
      read streams: {bytes: 8.00 B, streams: 1}
      read+write streams: {bytes: 0.00 B, streams: 0}
      write streams: {bytes: 8.00 B, streams: 1}
    ...
  measurements:
    L1:
      1:
        cores: [1, 2, 3, ...]
        results:
          copy: [36.15 GB/s, 72.32 GB/s, 108.48 GB/s, ...]
          ...
        size per core: [21.12 kB, 21.12 kB, 21.12 kB, ...]
        size per thread: [21.12 kB, 21.12 kB, 21.12 kB, ...]
        threads: [1, 2, 3, ...]
        threads per core: 1
        total size: [21.12 kB, 42.24 kB, 63.36 kB, ...]
    ...

\end{lstlisting}

\paragraph{Compute Architecture}
The first section is the execution architecture description (the actual order of elements does not matter in the YAML file format). This section describes the compute capabilities of the machine, such as clock speed, number of cores, or compiler flags to use for benchmarks. \texttt{micro-architecture} is the abbreviation for the Intel microarchitecture codename as used by IACA (e.g., \texttt{HSW} for Haswell), \texttt{overlapping-ports} are the execution ports corresponding to the overlapping portion in the ECM model as reported by IACA, \texttt{non-overlapping-ports} are all other ports as reported by IACA.

The machine description file with the benchmark section and partial information about the memory hierarchy and compute architecture can be generated automatically by the script \texttt{likwid\_bench\_auto.py}, which comes with Kerncraft. It employs \texttt{likwid-topology} and \texttt{likwid-bench} \cite{likwid}  to gather accurate information about the machine it is executed on.

\paragraph{Memory Hierarchy}
Each level of the memory hierarchy has an entry in the \texttt{memory hierarchy} section. \texttt{cores per group} is the number of physical cores that share one resource on this level (e.g., if every core has its own private cache, \texttt{cores per group} is 1). \texttt{threads per group} is the number of virtual threads that share one resource on this level. \texttt{groups} is the total number of resources of this type (e.g., an L1 cache) that exist on all sockets. \texttt{cycles per cacheline transfer} is only needed for caches, except for the last level cache (LLC). It denotes the number of cycles it takes to load one cache line from the adjacent ``lower'' (closer to main memory) cache. For the last level cache this number is calculated from the measured saturated memory bandwidth. 

The \texttt{cache per group} dictionary contains the cache description as required by pycachesim \cite{pycachesim}. \texttt{write\_back} makes sure that a modified cache line is transferred to the \texttt{store\_to} cache in case of its replacement. \texttt{write\_allocate} enforces a load of the cache line if some part of it is updated. The product of \texttt{sets}, \texttt{ways}, and \texttt{cl\_size} is the size of one cache resource in bytes.

\paragraph{Benchmarks}
Streaming benchmark results are required input for the Roof{}line model with all core counts and in all memory hierarchy levels. The ECM model only needs the measured saturated main memory bandwidth. In order to cover the whole memory hierarchy and typical effects and configurations, many tests are performed and their results stored in the machine description file. First, all benchmark kernels need to be specified in the \texttt{kernels} dictionary. For each kernel, \texttt{FLOPs per iteration} is the number of floating-point operations per iteration of the underlying kernel. \texttt{read streams} is the number of bytes and different streams read at each iteration. The ratio $\mathrm{\texttt{bytes}}/\mathrm{\texttt{streams}}$ is the size of one element in the processed array. \texttt{read+write streams} are accesses that are both read and written to (e.g., \texttt{a} in \texttt{a[i] = a[i] + 1}). \texttt{write streams} complements \texttt{read streams}. The differentiation into these three metrics is important to handle write-allocate transfers correctly.

The benchmark results are then grouped into memory hierarchy levels and SMT threads. Each such block has the configuration per physical core, with measured bandwidth (without write-allocate), used memory size (total, per thread and per core), and the number of cores and threads used.

\subsection{Models}
\label{subsec:models}
The models offered in Kerncraft are: \texttt{Roof{}line}, \texttt{ECM}, \texttt{Layer Conditions}, and \texttt{Benchmark}. Although not all are, strictly speaking, performance models, each one allows some unique and valuable insight into the performance, or some aspect of expected behavior, of the kernel at hand.

\paragraph{Roofline}
The Roof{}line model is implemented in the two variants \texttt{Roofline} and \texttt{Roof\-line\-IACA}. The former counts \flops\ in the high level code and matches them to the \texttt{FLOPs per cycle} configuration in the machine description file. It also models the first level cache to register transfers using the corresponding measured bandwidth result. \texttt{RooflineIACA}, on the other hand, uses the IACA analysis to predict in-core or compute performance and first level cache to register throughput. This analysis will be explained in detail in the ECM section below.

Apart from the differences in the in-core and first level cache to registers bottlenecks, both variants use the same approach for analysis throughout the rest of the memory hierarchy: take the cache miss prediction (explained in Section~\ref{subsec:cacheprediction}) and predict the required data volume ($\beta_k$) coming out of each memory hierarchy level per iteration. Take these volumes and divide them by the measured achievable bandwidths ($B_k$) out of the corresponding hierarchy level, which yields a throughput time for that data amount ($T_k={\beta_k}/{B_k}$).
Out of the numerous benchmarks (as described in Section~\ref{subsec:machine_description}), Kerncraft tries to find the one matching the kernel under examination as closely as possible with regard to the number of read and write streams into memory.

If IACA is available and a supported Intel architecture is analyzed, the \texttt{Roof\-line\-IACA} model is to be preferred over the regular \texttt{Roofline} model, as it will yield a much better accuracy.

\paragraph{ECM}

Three versions of the Execution-Cache-Memory (ECM) model are supported: \texttt{ECM\-Data} (modeling only the first level cache to main memory data transfers), \texttt{ECMCPU} (modeling only the in-core performance and first level cache to registers) and \texttt{ECM} (combining the data and in-core predictions). \texttt{ECMPCPU} relies on a suitable compiler and IACA to be available, which is why the rest of the ECM model can be run separately.

\texttt{ECMData} uses either the layer conditions or cache simulation (both explained in Section~\ref{subsec:machine_description}) to predict the data volumes out of each memory hierarchy level. Then it applies the documented bandwidths for inter-cache transfers and the measured full-socket main memory bandwidth for the memory to LLC transfers. By taking the ratio of data volume and bandwidth, $T_\mathrm{L1L2}$, $T_\mathrm{L2L3}$, and $T_\mathrm{L3Mem}$ are calculated (on machines with three cache levels). The benchmark kernel used for the main memory bandwidth is chosen according to the read and write stream counts best matching the analyzed kernel.

\texttt{ECMCPU} requires that the kernel is analyzed by IACA, which in turn requires a compilable version of the kernel. The kernel code is therefore wrapped in a \verb.main. function that takes care of initializing all arrays and variables. Dummy function calls are inserted to prevent the compiler from removing seemingly useless data accesses. Once compiled to assembly language using appropriate optimizing flags, the innermost kernel loop is extracted and the unrolling factor is determined from it. Both are done using heuristics and may fail; if they do, interaction by the user is requested. Using the unrolling factor, the IACA predictions can be scaled to iterations in the high-level kernel code. IACA reports throughput cycle counts per port, which are then accumulated into $T_\mathrm{nOL}$ and $T_\mathrm{OL}$ based on the machine description configuration.

\paragraph{Layer Conditions}

To predict optimal blocking sizes, layer conditions can be formulated in an algebraic way and solved for block sizes. The details are explained in \cite{lc-web}, while the concept of layer conditions and our generic formulation is described in Section~\ref{subsec:layer-conditions}.

\paragraph{Benchmark}

To allow validation of the previously explained models, the benchmark model compiles and runs the code and measures performance. The code is prepared in basically the same way as for an IACA analysis, but arrays are initialized and LIKWID marker calls are inserted to enable precise measurements using hardware performance counters. The output of \verb.likwid-perfctr. is used to derive familiar metrics (\GFS, \MLUPS, etc.), which in turn are used for validations. It is important to note that this model must be executed on the same machine as the one in the machine description file passed to Kerncraft, otherwise results will not be conclusive.

\subsection{Cache Miss Prediction}
\label{subsec:cacheprediction}

One of the core capabilities of Kerncraft is the prediction of the origin of data within the memory hierarchy, which can currently be done via two methods: a partial cache simulation using pycachesim, or a layer condition analysis. Both prediction methods have their strong points and drawbacks. Cache simulation can capture some irregularities arising from the cache structure and implementation in hardware (such as associativity conflicts) and at the same time is more generic and versatile in terms of architectural features and the kernels it can be used for. Layer conditions, on the other hand, yield very clean and stable results without disturbance from hardware-specific issues. They can be evaluated very quickly and almost independently of the code and domain size, but they only work for least-recently-used (LRU) replacement policies and currently only handle sequential traversal patterns.

In summary, if the layer condition prediction can be applied to the kernel and architecture of interest, it is usually the better choice.

\subsubsection{Cache Simulation with pycachesim}
\label{subsec:pycachesim}

The open source pycachesim library is a spin-off from Kerncraft. It is designed to efficiently model all the common cache architectures found in Intel, AMD, and Nvidia products.\footnote{Kerncraft currently only supports Intel Xeon and Core architectures, but pycachesim has been developed with other architectures in mind.} The cache architecture is described in the machine description file and then modeled in pycachesim. It supports inclusive and exclusive caching, multiple replacement policies (LRU, RR, Random and FIFO) as well as victim caches. For the Intel architectures covered in this paper, inclusive write-back caches with LRU are assumed. The simulator, once initialized with the cache structure, gets passed accessed data locations (loads and stores), which are followed through the simulated memory hierarchy. It also keeps a statistic about accumulated load, store, hit, and miss counts. After a warm-up phase, the statistic is reset, data accesses from a precise number of loop iterations are passed to the simulator, and the updated statistic is read out. The gained information reflects the steady state behavior.

It is very important to align the end of the warm-up period with cache line boundaries, as well as with edges of the arrays to skip over boundary handling (e.g., loops that go from $2$ to $N-3$). If these cases are not considered, imprecise and oscillating performance predictions are likely.

\subsubsection{Layer Conditions}
\label{subsec:layer-conditions}

Another approach to predicting the cache traffic are the Layer Conditions \cite{Rivera:2000,stengel14}. In order to utilize them for our purposes, we have generalized and reformulated them to allow symbolic evaluation. The symbolic evaluation heavily relies on sympy \cite{sympy}, a computer algebra system for python.

The basis of layer conditions is the least-recently-used replacement policy, which (although typically not perfectly implemented in large, real caches) mimics observed behavior quite well. By taking the relative data access offsets and assuming sequential increments during the subsequent iterations, we can predict very precisely which access will hit or miss depending on given cache sizes.

For demonstration we assume a double precision 2D 5-point stencil on $\mathrm{M}\times\mathrm{N}$ arrays \texttt{a[M][N]} and \texttt{b[M][N]}, with accesses in the $j$th and $i$th iteration to  \texttt{a[j-1][i]},  \texttt{a[j][i-1]} \texttt{a[j][i+1]}, \texttt{a[j+1][i]} and \texttt{b[j][i]}. The inner loop index is \texttt{i}. Now we compute the offsets between all accesses after sorting them in increasing order (as already shown), e.g., \texttt{\&a[j][i-1] - \&a[j-1][i]} or $(N-1)$ elements. We store them in the list $L$ and insert, per array, another $\infty$, since we do not know the offsets between the arrays:
$$L=\{\underbrace{\infty}_{\text{first access}\atop\text{to \texttt{a}}}, \underbrace{N-1}_{\text{\texttt{\&a[j][i-1]}}\atop\text{\texttt{- \&a[j-1][i]}}}, \underbrace{2}_{\text{\texttt{\&a[j][i+1]}}\atop\text{\texttt{- \&a[j][i-1]}}}, \underbrace{N-1}_{\text{\texttt{\&a[j+1][i]}}\atop\text{\texttt{- \&a[j][i+1]}}}, \underbrace{\infty}_{\text{first access}\atop\text{to \texttt{b}}}\}$$
For each reuse distance $t$ in $L$ we can derive the required cache size $C_\mathrm{req}$, hits $C_\mathrm{hits}$, and misses $C_\mathrm{misses}$:
\begin{eqnarray*}
C_\mathrm{req} & =  & \sum(L_{\leq t}) + t*\mathrm{count}(L_{>t})\\
C_\mathrm{hits} & =  & \mathrm{count}(L_{\leq t})\\
C_\mathrm{misses} & = & \mathrm{count}(L_{>t})~~.
\end{eqnarray*}
Here, $L_\mathrm{condition}$ is a sublist of $L$ that contains only entries that fulfill the given condition (e.g., $L_{<t}$ contains all elements out of $L$ which are smaller than $t$).
Applying this method to the described kernel, we have the interesting case $t=N-1$, for which we get $C_\mathrm{req} = 2(N-1)+2+2(N-1) = 4N-2$ elements, or $32N-16$\,bytes, $C_\mathrm{hits} = 3$, and $C_\mathrm{misses} = 2$.

This means that if an LRU-based cache can hold more than  $32N-16\,\mathrm{bytes}$, three hits will be observed in each iteration and two misses will need to be passed to the next level in the memory hierarchy, which is to leading order exactly the result from a manual LC analysis (where the 16\,bytes are typically neglected so that four layers, i.e., rows, must fit into the cache). Since caches in modern CPUs do not operate on bytes but on cache lines, the computed hits and misses are averaged. Once a cache line was loaded due to a miss, subsequent accesses will be hits, which averages out to the misses and hits per iteration yielded by the layer condition analysis.

\subsection{Underlying In-Core Execution Prediction}
\label{subsec:in-core}

To predict the in-core execution behavior, we employ the Intel Architecture Core Analyzer (IACA) \cite{iacaweb}, which predicts the throughput and latency for a sequence of assembly instructions under the assumption that all loads can be served by the first level cache. IACA presupposes steady-state execution, i.e., the loop body is assumed to be executed often enough to amortize any start-up effects.

Kerncraft operates on high level C code, which can not be analyzed by IACA directly. Therefore it first needs to be transformed into a compilable version by wrapping the kernel in a \verb.main. function. It is then passed through a compiler and converted to assembly. The assembly sequence of the inner loop body needs to be marked to be recognized by IACA. The marked assembly is then fed into the assembler to produce an object file as input to IACA. IACA reports the throughput and latency analysis itemized by execution ports.
We are interested in the overall and load-related throughput and latency. Which execution ports are associated with loads is defined in the machine description file (see Section~\ref{subsec:machine_description} above).
The compiler might have unrolled the inner-most loop a number of times (e.g., to allow vectorization), so this factor needs to be extracted from the assembly to scale the IACA results to a single high-level kernel code loop iteration. The IACA output is parsed and the data is presented by Kerncraft as part of the analysis.

\section{Kerncraft Usage}
\label{sec:validation}

Kerncraft guides performance engineering efforts by allowing developers to predict and validate performance. In the following sections we will use an instructive example to demonstrate the single-core performance prediction, the scaling from single-core to the full socket, and the analytic layer conditions.
The analysis will be based on the long-range 3D kernel (3d-long-range) in Listing~\ref{lst:3d-long-range}.
Predictions and measurements will be done for the Intel Ivy Bridge EP (IVY) microarchitecture. The details of the machine are described in Table~\ref{tbl:machines}.

\begin{lstlisting}[language=C,float=tbp,label={lst:3d-long-range},caption={Kernel code for a three dimensional long-range star stencil with constant coefficients.}]
double U[M][N][N];
double V[M][N][N];
double ROC[M][N][N];
double c0, c1, c2, c3, c4, lap;

for(int k=4; k < M-4; k++) {
    for(int j=4; j < N-4; j++) {
        for(int i=4; i < N-4; i++) {
            lap = c0 * V[k][j][i]
                + c1 * ( V[ k ][ j ][i+1] + V[ k ][ j ][i-1])
                + c1 * ( V[ k ][j+1][ i ] + V[ k ][j-1][ i ])
                + c1 * ( V[k+1][ j ][ i ] + V[k-1][ j ][ i ])
                + c2 * ( V[ k ][ j ][i+2] + V[ k ][ j ][i-2])
                + c2 * ( V[ k ][j+2][ i ] + V[ k ][j-2][ i ])
                + c2 * ( V[k+2][ j ][ i ] + V[k-2][ j ][ i ])
                + c3 * ( V[ k ][ j ][i+3] + V[ k ][ j ][i-3])
                + c3 * ( V[ k ][j+3][ i ] + V[ k ][j-3][ i ])
                + c3 * ( V[k+3][ j ][ i ] + V[k-3][ j ][ i ])
                + c4 * ( V[ k ][ j ][i+4] + V[ k ][ j ][i-4])
                + c4 * ( V[ k ][j+4][ i ] + V[ k ][j-4][ i ])
                + c4 * ( V[k+4][ j ][ i ] + V[k-4][ j ][ i ]);
            U[k][j][i] = 2.f * V[k][j][i] - U[k][j][i]
                       + ROC[k][j][i] * lap;
}}}
\end{lstlisting}

\begin{table}[tbp]
    \centering\renewcommand{\arraystretch}{1.1}
    \label{tbl:machines}
    \caption{Technical data of the Ivy Bridge-based node used for the long-range stencil case study.}
\begin{tabular}{r||c}
    Microarchitecture & Ivy Bridge EP \\
    \hline
    Abbreviation      & IVY           \\
    \hline
    \hline
    Model Name & E5-2690v2 \\
    \hline
    Clock (fixed, no turbo) & 3.0 GHz\\
    \hline
    Cores per socket & 10 \\
    \hline\hline
    Cacheline size & $64\,\mathrm{B}$ \\
    \hline
    Theoretical L1-L2 bandwidth & $0.5\,\mathrm{CL/cy}$ \\
    \hline
    Theoretical L2-L3 bandwidth per core & $0.5\,\mathrm{CL/cy}$ \\
    \hline
    Achievable single-socket memory &  \multirow{2}{*}{$47.2\,\mathrm{GB/s}$ (7 cores)}\\
    bandwidth (copy kernel) &  \\
    \hline
    \hline
    Compiler version & Intel ICC 16.0.3 \\
    \hline
    IACA version & 2.1 \\
    \hline
    Kerncraft version & 0.4.3 \\
\end{tabular}
\end{table}

\subsection{Single-Core Performance}
\label{subsec:single-core}
\begin{figure}[tbp]
\sidecaption
\includegraphics[width=\textwidth]{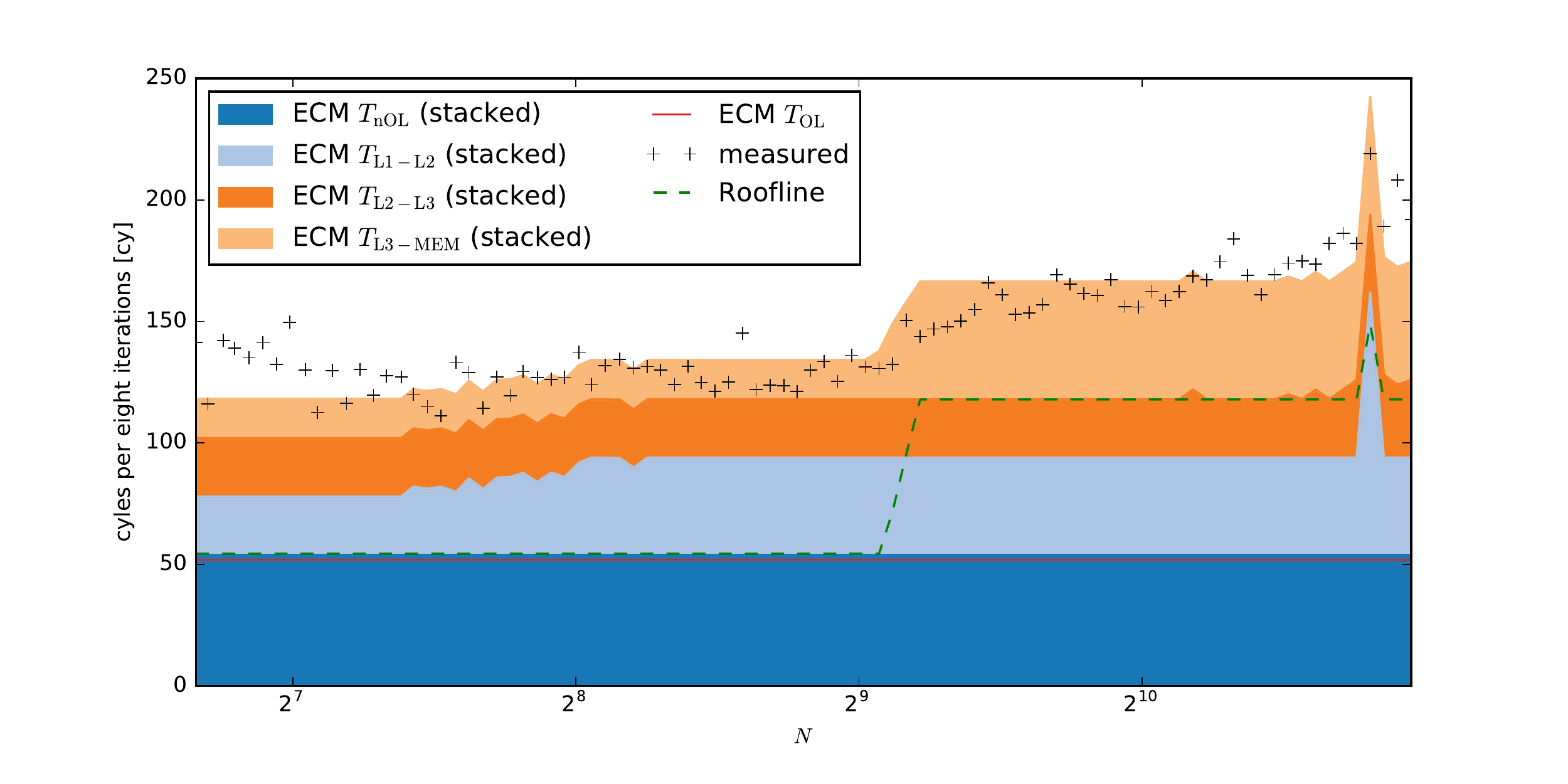}
\caption{Single-core parameter sweep of the long-range stencil for $N=100$ to $N=2000$ with $M$ chosen such that the working set will never fit into any cache and needs to be loaded from main memory.}
\label{fig:single-core}
\end{figure}

Using Kerncraft for a single-core performance analysis involves choosing an overall prediction model (ECM or Roof{}line) and a cache predictor model (pycachesim simulation [SIM] or layer conditions [LC]). An example using \verb.RooflineIACA., \verb.ECM., and \verb.SIM. is shown in Listing~\ref{lst:output_ecm_rl}.
It is easy to do parameter studies via simple scripting, and scanning a range of problem sizes often leads to valuable insights. Running this analysis from $N=100$ to $N=2000$, we can see the effect of the inner dimension increasing and visualize it in Figure~\ref{fig:single-core}.
\begin{lstlisting}[float=tbp,language=shell,label={lst:output_ecm_rl},caption={Excerpt from the kerncraft CLI (reformatted for brevity) for the analysis of the long-range stencil}]
$ kerncraft -p ECM -p RooflineIACA --cache-predictor=SIM \
            3d-long-range.c -m IVY.yaml -D M 130 -D N 1015;
=========================== kerncraft ===========================
3d-long-range-stencil.c                               -m IVY.yaml
-D M 130 -D N 1015
----------------------------- ECM -------------------------------
{ 52.0 || 54.0 | 40.0 | 24.0 | 48.5 } cy/CL
{ 54.0 \ 94.0 \ 118.0 \ 166.5 } cy/CL
saturating at 4 cores

------------------------- RooflineIACA --------------------------
Bottlenecks:
 level | a. intensity | performance   | bandwidth  | bw kernel
-------+--------------+---------------+------------+----------
   CPU |              | 18.22 GFLOP/s |            |
    L2 |  0.26 FLOP/B | 17.52 GFLOP/s | 68.37 GB/s | copy
    L3 |  0.43 FLOP/B | 16.57 GFLOP/s | 38.79 GB/s | copy
   MEM |  0.43 FLOP/B |  7.65 GFLOP/s | 17.91 GB/s | copy

Cache or mem bound with 1 core(s)
7.65 GFLOP/s due to MEM bottleneck (bw with from copy benchmark)
Arithmetic Intensity: 0.43 FLOP/B
\end{lstlisting}

\begin{figure}[!h]
\sidecaption
\includegraphics[width=\textwidth]{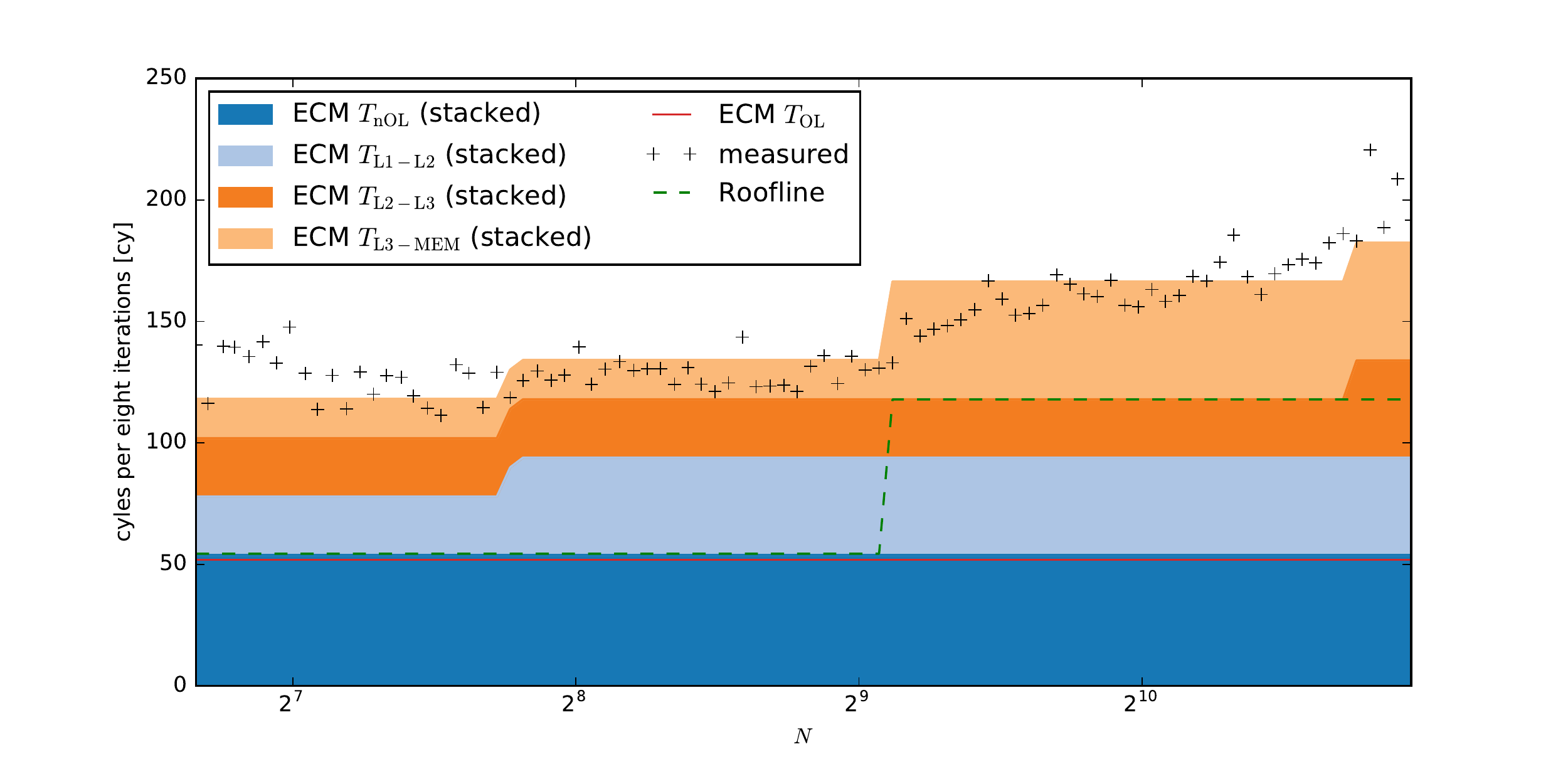}
\caption{Single-core parameter sweep, with layer condition cache prediction, of the long-rang stencil for $N=100$ to $N=2000$ with $M$ chosen such that the data will never fit into any cache and needs to be loaded from main memory.}
\label{fig:single-core_LC}
\end{figure}

The ECM prediction (stacked areas from $T_\mathrm{nOL} + T_\mathrm{L1-L2} + T_\mathrm{L2-L3} + T_\mathrm{L3-MEM}$) follows the trend of the measured throughput (black plus signs). The Roof{}line prediction (green dashed line) is generally too optimistic due to the evenly distributed runtime contribution from multiple memory hierarchy levels, which is not correctly modeled in this particular case. The cache simulator, taking the associativity of all cache levels into account, correctly identifies L1 thrashing and a corresponding runtime increase near $N=1792=7\cdot 256$. The corresponding increase in traffic between L1 and L2 of more than 50\% can be shown using performance counter measurements.
Many more such ``pathological'' sizes exist, of course, but the size range was not scanned with a step size of one.
In Figure~\ref{fig:single-core_LC} the same parameter study was done with the LC predictor. Since it knows nothing about cache organization, the prediction is much smoother.
%

\subsection{Single-Socket Scaling and Saturation Point}
\label{subsec:scaling}

For multi-core scaling the ECM model assumes perfect scalability until a shared
bandwidth bottleneck (usually the main memory bandwidth) is hit. It
thus predicts the number of cores where the loop performance ceases to
scale: 
$$
n_\mathrm s=\frac{T_\mathrm{ECM,Mem}}{T_\mathrm{L3-Mem}}~~.
$$
By default, Kerncraft reports the saturation point in the ECM model, as seen in Listing~\ref{lst:output_ecm_rl}. The default report assumes that the total cache size and cache bandwidth scales with the number of cores. This is mostly true on current Intel microarchitectures, but not for the last level cache (L3) size, which is shared among all cores in a socket. To also take that change of cache sizes into account, Kerncraft can be run with the {\tt--cores} argument.
In the case presented in Listing~\ref{lst:output_ecm_rl}, a reduction of the L3 cache size by a factor of four (for 4 cores) does not change the predicted results, since no layer condition changes.

\begin{figure}[bp]
\sidecaption
\includegraphics[width=\textwidth]{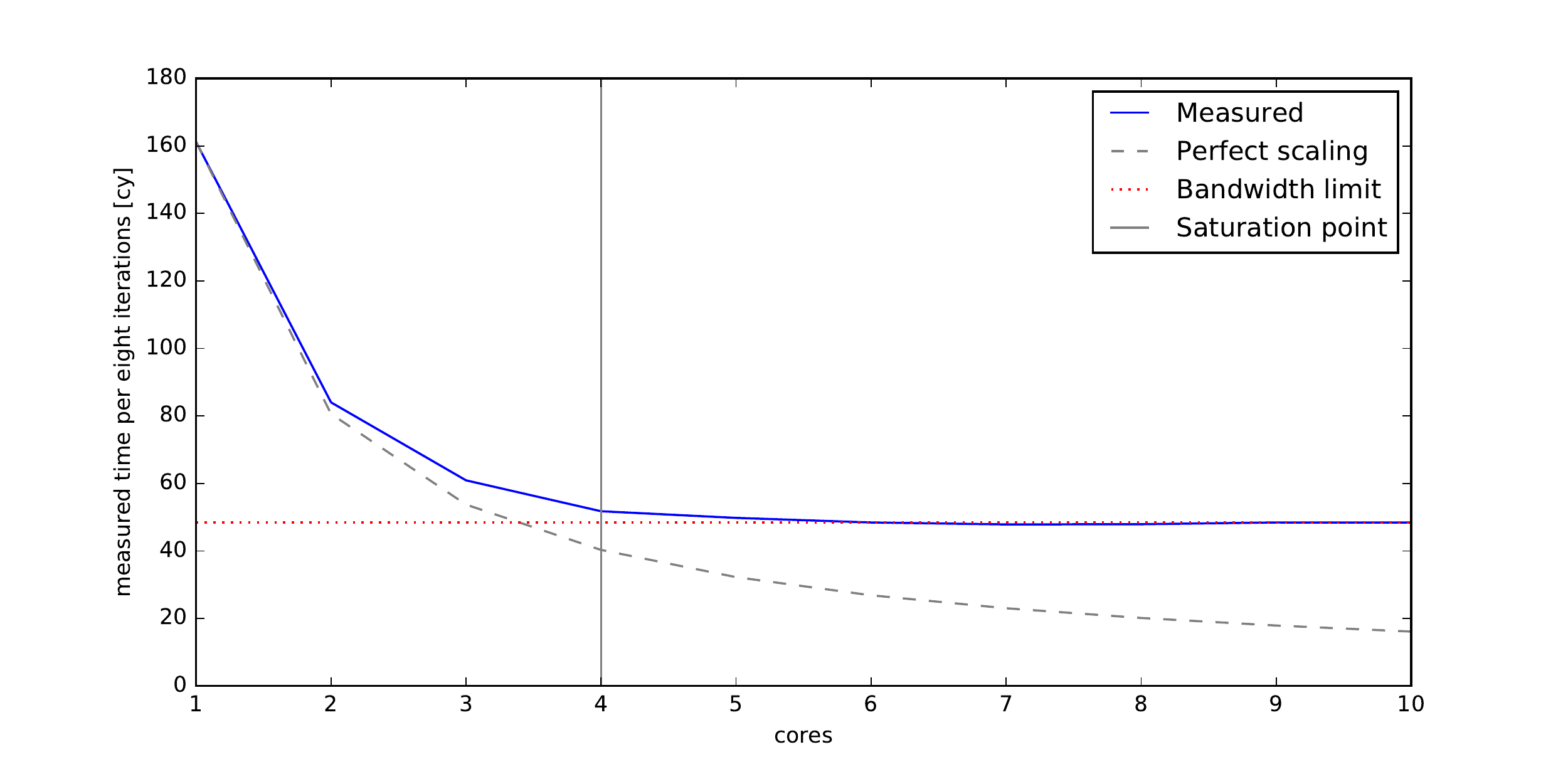}
\caption{Single-socket strong scaling of the long-range stencil for $N=1015$ and $M=132$ with all cores on same socket. The vertical line denotes the predicted saturation point. The horizontal line is the minumum runtime as given by the saturated memory bandwidth.}
\label{fig:single-socket}
\end{figure}

To perform the single-socket scaling  we added OpenMP pragmas to the outer loop in the code and ran with the same problem size as seen in Listing~\ref{lst:output_ecm_rl} (strong scaling). The result can be seen in Figure~\ref{fig:single-socket}: By increasing the number of cores up to the predicted saturation point (four cores), we expect perfect scaling (dashed gray line), and constant performance beyond (dotted line). The scaling model fits the observations very well except right before the saturation point, which is a known weakness of the ECM model with data-bound kernels \cite{stengel14}. 

\subsection{Layer Conditions}
\label{subsec:loop-tiling}

Layer conditions enable a much more efficient cache behavior prediction without extensive parameter studies through the simulator or benchmarks. As explained in Section~\ref{subsec:layer-conditions}, they are evaluated analytically and yield a prediction for transition points from one cache state to another. Kerncraft generally employs analytic LCs when using the option {\tt--cache-predictor=LC}, but it can also output the derived transition points as shown in Listing~\ref{lst:kerncraft_lc}. The predicted transition in L3 from the 3D to the 2D layer condition at $N = 546$ is also clearly visible in Figures~\ref{fig:single-core} and \ref{fig:single-core_LC}.

\begin{lstlisting}[float=tbp,language=shell,label={lst:kerncraft_lc},caption={Excerpt from the kerncraft CLI (reformatted for brevity) showing LC transition points from the analysis of the long-range stencil}]
$ kerncraft -p LC 3d-long-range.c -m IVY.yaml -D M 130 -D N 1015;
=========================== kerncraft ===========================
3d-long-range-stencil.c                               -m IVY.yaml
-D M 130 -D N 1015
------------------------------ LC -------------------------------
2D Layer-Condition:
L1: N <= 216
L2: N <= 1725
L3: N <= 172463
3D Layer-Condition:
L1: N <= 19
L2: N <= 55
L3: N <= 546
\end{lstlisting}

\section{Future Work}
\label{sec:future}
Development on Kerncraft will continue and strive to enhance usability and portability and to allow support of a broader range of kernels and architectures.
One of the major obstacles to supporting  non-Intel CPUs is IACA, which is closed-source and only supports Intel microarchitectures. It is our goal to develop a model and tool which will be suitable for predictions on other architectures. 
In the near future we will also integrate our layer condition model with the LLVM-Polly project \cite{polly:2012}.
This will allow the Polyhedral model to automatically choose cache-efficient tiling sizes without user interaction.

As with all of our tools and libraries (Kerncraft, LIKWID \cite{likwid}, GHOST \cite{ghost}, and the soon-to-be-published fault-tolerance package CRAFT), future work will be released under open source licenses and we will support and encourage other projects to build upon them.

\begin{acknowledgement}
This work was in part funded by the German Academic Exchange Service's (DAAD) FITweltweit program and the Federal Ministry of Education and Research (BMBF) SKAMPY grant. 
\end{acknowledgement}

\bibliographystyle{spmpsci}
\bibliography{references}

\begin{thebibliography}{10}
\providecommand{\url}[1]{{#1}}
\providecommand{\urlprefix}{URL }
\expandafter\ifx\csname urlstyle\endcsname\relax
  \providecommand{\doi}[1]{DOI~\discretionary{}{}{}#1}\else
  \providecommand{\doi}{DOI~\discretionary{}{}{}\begingroup
  \urlstyle{rm}\Url}\fi

\bibitem{kerncraft-github}
Kerncraft toolkit.
\newblock \url{https://github.com/RRZE-HPC/kerncraft}

\bibitem{djoudi2005maqao}
Djoudi, L., Barthou, D., Carribault, P., Lemuet, C., Acquaviva, J.T., Jalby,
  W., et~al.: {MAQAO}: Modular assembler quality analyzer and optimizer for
  itanium 2.
\newblock In: The 4th Workshop on {EPIC} architectures and compiler technology,
  San Jose (2005).
\newblock \url{http://www.prism.uvsq.fr/users/bad/Research/ps/maqao.pdf}

\bibitem{yaml}
Evans, C., Ingerson, B., Ben-Kiki, O.: {YAML Ain't Markup Language}.
\newblock \url{http://yaml.org}

\bibitem{polly:2012}
Grosser, T., Groesslinger, A., Lengauer, C.: Polly -- performing polyhedral
  optimizations on a low-level intermediate representation.
\newblock Parallel Processing Letters \textbf{22}(04), 1250,010 (2012).
\newblock \doi{10.1142/S0129626412500107}

\bibitem{lc-web}
Hammer, J.: Layer conditions.
\newblock \urlprefix\url{https://rrze-hpc.github.io/layer-condition/}

\bibitem{pycachesim}
Hammer, J.: pycachesim -- a single-core cache hierarchy simulator written in
  python.
\newblock \urlprefix\url{https://github.com/RRZE-HPC/pycachesim}

\bibitem{hammer15kerncraft}
Hammer, J., Hager, G., Eitzinger, J., Wellein, G.: Automatic loop kernel
  analysis and performance modeling with kerncraft.
\newblock In: Proceedings of the 6th International Workshop on Performance
  Modeling, Benchmarking, and Simulation of High Performance Computing Systems,
  PMBS '15, pp. 4:1--4:11. ACM, New York, NY, USA (2015).
\newblock \doi{10.1145/2832087.2832092}

\bibitem{hockney89}
Hockney, R.W., Curington, I.J.: {$f_{1/2}$}: A parameter to characterize memory
  and communication bottlenecks.
\newblock Parallel Computing \textbf{10}(3), 277--286 (1989).
\newblock \doi{10.1016/0167-8191(89)90100-2}

\bibitem{Hofmann:2016-Kahan}
Hofmann, J., Fey, D., Riedmann, M., Eitzinger, J., Hager, G., Wellein, G.:
  Performance analysis of the {Kahan}-enhanced scalar product on current
  multi-core and many-core processors.
\newblock Concurrency and Computation: Practice and Experience pp. n/a--n/a
  (2016).
\newblock \doi{10.1002/cpe.3921}

\bibitem{iacaweb}
{I}ntel {A}rchitecture {C}ode {A}nalyzer.
\newblock
  \urlprefix\url{https://software.intel.com/en-us/articles/intel-architecture-code-analyzer}.
\newblock
  \url{https://software.intel.com/en-us/articles/intel-architecture-code-analyzer}

\bibitem{ISO:C99}
ISO: {ISO C Standard 1999}.
\newblock Tech. rep. (1999).
\newblock
  \urlprefix\url{http://www.open-std.org/jtc1/sc22/wg14/www/docs/n1124.pdf}.
\newblock ISO/IEC 9899:1999 draft

\bibitem{ghost}
Kreutzer, M., Thies, J., R{\"o}hrig-Z{\"o}llner, M., Pieper, A., Shahzad, F.,
  Galgon, M., Basermann, A., Fehske, H., Hager, G., Wellein, G.: {GHOST}:
  Building blocks for high performance sparse linear algebra on heterogeneous
  systems.
\newblock International Journal of Parallel Programming pp. 1--27 (2016).
\newblock \doi{10.1007/s10766-016-0464-z}

\bibitem{pbounds2010}
Krishna~Narayanan, S.H., Norris, B., Hovland, P.D.: Generating performance
  bounds from source code.
\newblock In: Parallel Processing Workshops (ICPPW), 2010 39th International
  Conference on, pp. 197--206 (2010).
\newblock \doi{10.1109/ICPPW.2010.37}

\bibitem{rmt15}
Lo, Y., Williams, S., Van~Straalen, B., Ligocki, T., Cordery, M., Wright, N.,
  Hall, M., Oliker, L.: Roof\/line model toolkit: A practical tool for
  architectural and program analysis.
\newblock In: S.A. Jarvis, S.A. Wright, S.D. Hammond (eds.) High Performance
  Computing Systems. Performance Modeling, Benchmarking, and Simulation,
  \emph{Lecture Notes in Computer Science}, vol. 8966, pp. 129--148. Springer
  International Publishing (2015).
\newblock \doi{10.1007/978-3-319-17248-4\_7}.
\newblock DOI:~10.1007/978-3-319-17248-4\_7

\bibitem{STREAM}
McCalpin, J.D.: {STREAM}: Sustainable memory bandwidth in high performance
  computers.
\newblock Tech. rep., University of Virginia, Charlottesville, VA (1991-2007).
\newblock \urlprefix\url{http://www.cs.virginia.edu/stream/}.
\newblock A continually updated technical report

\bibitem{Rivera:2000}
Rivera, G., Tseng, C.W.: Tiling optimizations for {3D} scientific computations.
\newblock In: Supercomputing, ACM/IEEE 2000 Conference, pp. 32--32 (2000).
\newblock \doi{10.1109/SC.2000.10015}

\bibitem{stengel14}
Stengel, H., Treibig, J., Hager, G., Wellein, G.: Quantifying performance
  bottlenecks of stencil computations using the execution-cache-memory model.
\newblock In: Proceedings of the 29th ACM International Conference on
  Supercomputing, ICS '15, pp. 207--216. ACM, New York, NY, USA (2015).
\newblock \doi{10.1145/2751205.2751240}

\bibitem{sympy}
{SymPy Development Team}: SymPy: Python library for symbolic mathematics
  (2016).
\newblock \urlprefix\url{http://www.sympy.org}

\bibitem{likwid}
Treibig, J., Hager, G., Wellein, G.: Likwid: A lightweight performance-oriented
  tool suite for x86 multicore environments.
\newblock In: Proceedings of PSTI2010, the First International Workshop on
  Parallel Software Tools and Tool Infrastructures. San Diego CA (2010)

\bibitem{Unat01052015}
Unat, D., Chan, C., Zhang, W., Williams, S., Bachan, J., Bell, J., Shalf, J.:
  {ExaSAT}: An exascale co-design tool for performance modeling.
\newblock International Journal of High Performance Computing Applications
  \textbf{29}(2), 209--232 (2015).
\newblock \doi{10.1177/1094342014568690}.
\newblock DOI:~10.1177/1094342014568690

\bibitem{roofline:2009}
Williams, S., Waterman, A., Patterson, D.: Roof\/line: An insightful visual
  performance model for multicore architectures.
\newblock Commun. ACM \textbf{52}(4), 65--76 (2009).
\newblock \doi{10.1145/1498765.1498785}

\bibitem{Wittmann:2016}
Wittmann, M., Hager, G., Zeiser, T., Treibig, J., Wellein, G.: Chip-level and
  multi-node analysis of energy-optimized lattice {Boltzmann} {CFD}
  simulations.
\newblock Concurrency and Computation: Practice and Experience \textbf{28}(7),
  2295--2315 (2016).
\newblock \doi{10.1002/cpe.3489}

\end{thebibliography}
\end{document}